\documentclass[%
 aip,
 amsmath,amssymb,
 reprint,
]{revtex4-1}

\usepackage{graphicx}
\graphicspath{{Graphs/}}
\usepackage{dcolumn}
\usepackage{bm}
\usepackage[utf8]{inputenc}
\usepackage[T1]{fontenc}
\usepackage{mathptmx}
\usepackage{etoolbox}
\usepackage[dvipsnames]{xcolor}
\usepackage[colorlinks=true, urlcolor=blue, linkcolor=blue, citecolor=blue]{hyperref}

\makeatletter
\def\@email#1#2{%
 \endgroup
 \patchcmd{\titleblock@produce}
  {\frontmatter@RRAPformat}
  {\frontmatter@RRAPformat{\produce@RRAP{*#1\href{mailto:#2}{#2}}}\frontmatter@RRAPformat}
  {}{}
}%
\makeatother
\begin{document}


\title{Scattering anisotropy in HgTe\,(013) quantum well}

\author{D.\,A.\,Khudaiberdiev}
 \email{KhudaiberdievDA@gmail.com}
\affiliation{Rzhanov Institute of Semiconductor Physics, Novosibirsk 630090, Russia}
\affiliation{Institute of Solid State Physics, Vienna University of Technology, 1040 Vienna, Austria}
\author{M.\,L.\,Savchenko}
\affiliation{Rzhanov Institute of Semiconductor Physics, Novosibirsk 630090, Russia}
\affiliation{Institute of Solid State Physics, Vienna University of Technology, 1040 Vienna, Austria}
\author{D.\,A.\,Kozlov}
\affiliation{Rzhanov Institute of Semiconductor Physics, Novosibirsk 630090, Russia}
\affiliation{Experimental and Applied Physics, University of Regensburg, D-93040 Regensburg, Germany}
\author{N.\,N.\,Mikhailov}
\author{Z.\,D.\,Kvon}
\affiliation{Rzhanov Institute of Semiconductor Physics, Novosibirsk 630090, Russia}
\affiliation{Novosibirsk State University, Novosibirsk 630090, Russia}

\date{\today}

\begin{abstract}

We report on a detailed experimental study of the electron transport anisotropy in HgTe (013) quantum well of 22\,nm width in the directions $[100]$ and $[03\overline{1}]$ as the electron density function $n$.
The anisotropy is absent at the minimal electron density near a charge neutrality point.
The anisotropy increases with the increase of $n$ and reaches about 10$\%$ when the Fermi level is within the first subband H1.
There is a sharp increase of the anisotropy (up to 60$\%$) when the Fermi level reaches the second subband E2. 
We conclude that the first effect is due to the small intra-subband  anisotropic interface roughness scattering, 
and the second one is due to the strongly anisotropic inter-subband roughness scattering,
but the microscopical reason of such a strong change in the anisotropy remains unknown.

\end{abstract}

\maketitle

Experimental studies of HgTe-based quantum wells (QW) are  among the most actively developing directions of research in the physics of low-dimensional electron systems. 
Due to the strong relativistic effects in these QWs, different kinds of two-dimensional (2D) systems can be realized from them.
Films with a thickness lower than critical (6.5\,nm) are ordinary 2D insulators when the Fermi energy lies inside the band gap.

When the thickness is higher than critical, the hole-like subband H1 lies above the electron-like subband E1. 
Thus the system becomes a two-dimensional topological insulator\cite{Konig2007,Nowack2013,Konig2013,Dantscher2017}.
At a critical thickness Dirac fermion system occurs\cite{Buttner2011,Kvon2011,Kozlov2012,Gusev2017}. 
At higher thicknesses (starting with 14\,nm) H2 and H1 subbands overlap creating a 2D semimetal state (while the E1 subband lies deep in the valence band)\cite{Kvon2008,Olshanetsky2009,Minkov2013,Knap2014,Kononov2016}.

The highest quality HgTe QWs are grown by molecular-beam epitaxy in nonsingular directions, such as (013) and (112), with the electron mobility reaching the values of about $10^{6}$\,cm$^2$/Vs.
The reason of using such directions is to increase the dissociation rate of Te$_2$ molecules, to prevent crystallisation of clean Te, that is preferred near surface steps~\cite{Kvon2009b}.
In different 2D electron systems (based on the inverted GaAs/AlAs interface\cite{Markus1994}, InAs\cite{Le2021}, InSb\cite{Ball2002} and Ge\cite{Hassan2014} QWs) anisotropic transport caused by the anisotropic scattering on surface roughness was studied. In those works, even with the (001) growth direction, the scattering on roughness has a preferred direction $[110]$ which leads to the mobility anisotropy. 
Thus, HgTe QWs with the (013) nonsingular growth direction are even more expected to have the scattering anisotropy. 
We note that such anisotropy in the HgTe\,(013) QW was observed  before\cite{Minkov2013}, but the non-inverted 5\,nm-thick low electron mobility system was studied there.

In this paper we study the electron transport anisotropy in 22-nm HgTe QWs with the inverted spectrum, such that the first two conduction subbands are H1 and E2. 
We report on the transport anisotropy caused by the anisotropic roughness scattering when the Fermi level crosses only an H1 subband, and a non-trivial jump of anisotropy when the Fermi level reaches the second subband E2 caused by a strong anisotropic intra-subband scattering.

\begin{figure}[t]
        \includegraphics[width=1\columnwidth]{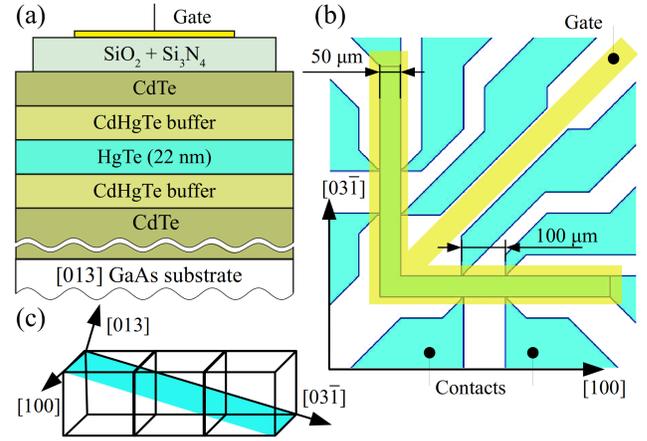}
        \caption{\label{fig:Sample} 
        (a)~Schematic cross-section of the heterostructures studied. 
        The HgTe 22\,nm film is located between Cd$_{0.65}$Hg$_{0.35}$Te buffer layers. 
        (b)~Schematic top view of the $L$-shaped mesa-structure with [100] and $[03\overline{1}]$ arm orientations. 
        (c)~Crystallographic directions in the system.
        }
    \end{figure}

The studied QWs are grown by molecular beam epitaxy on a virtual CdTe (013) substrate  (Fig.~\ref{fig:Sample}\,(a)) having a 0.3\% larger lattice constant than HgTe and causing tensile strain\cite{Brune2011,Savchenko2019}.
The system is equipped with the Ti/Au metallic gate which is located above 100\,nm + 200\,nm SiO$_{2}$/Si$_{3}$N$_{4}$ insulator layers.
Several $L$-shaped Hall bar devices were prepared. 
They gave quantitatively similar results. 
Thus, we present data from one device.
The $L$-shaped arm orientations are symmetric [100] and non-symmetric $[03\overline{1}]$~(Fig.~\ref{fig:Sample}\,(b)).
The symmetry-based names are illustrated in panel~(c), where the studied crystallographic directions are schematically presented.
The current channel width is 50\,$\mathrm{\mu}$m, the distance between the
pairs of longitudinal potential probes is 100\,$\mathrm{\mu}$m.
All the measurements were carried out using a standard lock-in technique with about 12\,Hz frequency and the current (0.01 -- 1)\,$\mathrm{\mu}$A applied in a perpendicular magnetic field $B$ up to 1.5\,T.
All the presented results were obtained at temperature 0.2\,K.

\begin{figure}[t]
        \includegraphics[width=0.5\textwidth]{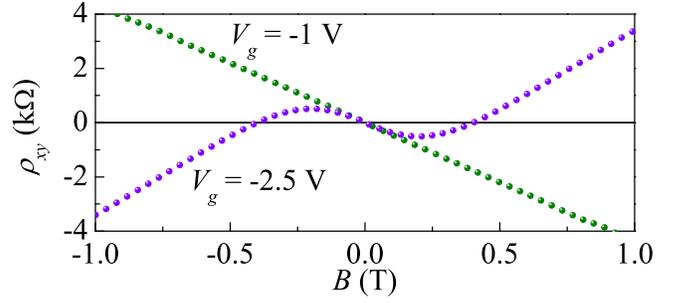}
        \caption{\label{fig:Rxy} 
        $\rho_{xy}(B)$ at $V_g = -1$ V (olive) and $V_g = -2.5$ (violet). 
        The nonlinear Hall resistance measured at $-2.5\,$V suggests the existence of holes and electrons at this $V_g$.
        }
    \end{figure}

In Fig.~\ref{fig:Rxx,Ryy_Fridge}\,(a) is the main result of our work: the gate voltage dependencies of the longitudinal resistance in directions [100] and $[03\overline{1}]$.
    \begin{figure}
        \includegraphics[width=0.5\textwidth]{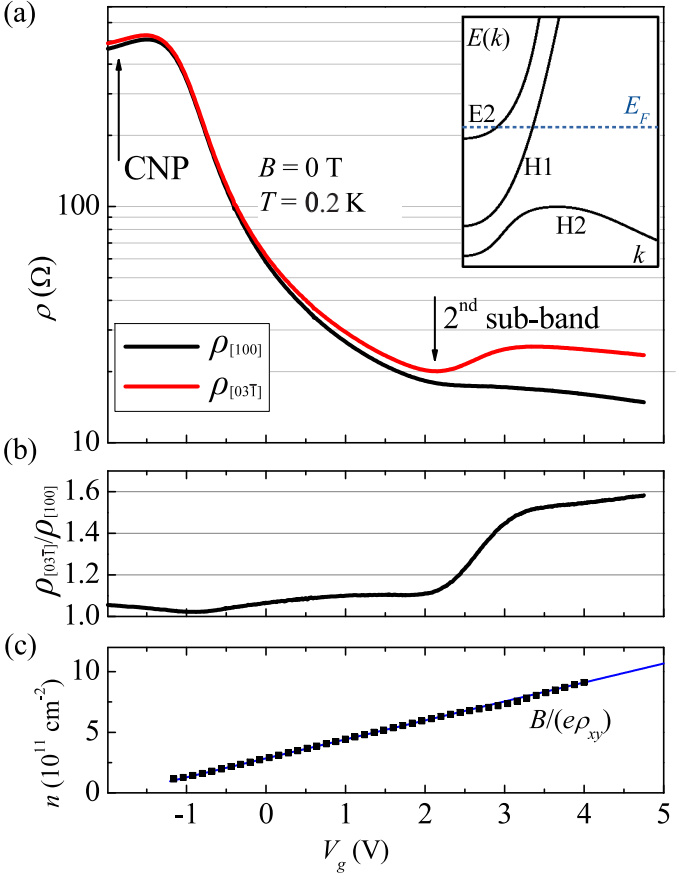}
        \caption{\label{fig:Rxx,Ryy_Fridge} 
        (a)~The gate voltage dependence of the longitudinal resistance in different directions: [100] (black) and $[03\overline{1}]$ (red). 
        (b)~The gate voltage dependence of the resistance anisotropy $\rho_{[03\overline{1}]}/\rho_{[100]}$.
        (c)~The gate voltage dependence of the electron density~(symbols) that was determined from the Hall effect measurements 
        at $B$ = 0.5\,T.}
    \end{figure}
Qualitatively, one can see the ordinary for (18-22)\,nm HgTe QWs gate voltage dependencies of resistance in both orientations: they have a maximum near the charge neutrality point (CNP), and there is a sharp decrease of resistances as the gate voltage increases. 
In the vicinity of CNP, the resistance behaviour is slightly anisotropic (see the gate voltage dependence of the resistance anisotropy ratio $\rho_{[03\overline{1}]}/\rho_{[100]}$ in Fig.~\ref{fig:Rxx,Ryy_Fridge}\,(b)) due to the hole mass anisotropy\cite{Minkov2017,Gospodaric2021}. 
The presence of holes is seen in~Fig.~\ref{fig:Rxy} as a non-linear
$\rho_{xy}(B)$ dependence measured at $V_g = -2.5$\,V~(violet symbols). 
We do not discuss this region in the current paper.
There are no holes already at $V_g = -1$\,V (see linear $\rho_{xy}(B)$ in~Fig.\,\ref{fig:Rxy}, green symbols), and the transport is fully isotropic here.
At higher gate voltages the difference between $\rho_{[03\overline{1}]}$ and $\rho_{[100]}$ gradually increases and their ratio saturates at around $\rho_{[03\overline{1}]}/\rho_{[100]} \approx 1.1$.
Surprisingly, at $V_{g} = 2$\,V the resistance along the direction $[03\overline{1}]$ starts to increase, while $\rho_{[100]}$ continues to 
decrease. 
So, their ratio $\rho_{[03\overline{1}]}/\rho_{[100]}$ jumps from 1.1 to 1.5 value.
And there is again only a moderate increase of the anisotropy ratio at higher gate voltages.

Such unexpected behaviour with the sharp increase of the anisotropy ratio at $V_g \approx$ (2 -- 3)\,V should come from a new scattering mechanism that starts in this gate voltage range, and that has a strong orientation dependence. 
But, before discussing the origin of such anisotropy dependence, let us analyze other transport properties of the system.

\begin{figure*}    
    \includegraphics[width=0.8\textwidth]{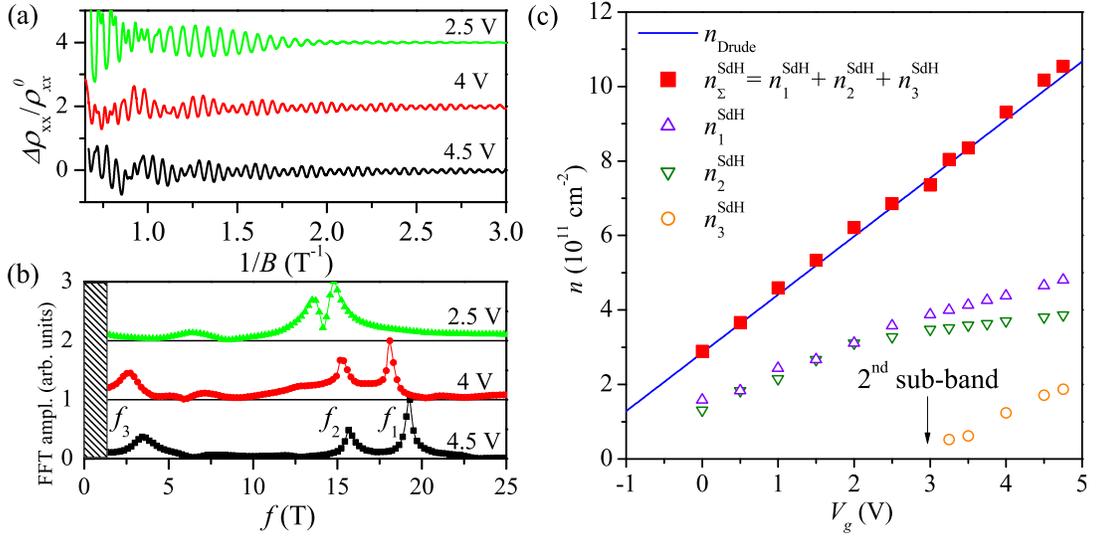}
    \caption{\label{fig:Shub_and_FFT_diffVg} 
        (a)~
        \textcolor{black}{Normalized} 
        resistance oscillations 
        $\Delta\rho_{xx}/\rho_{xx}^0 = (\rho_{xx}-\langle\rho_{xx}\rangle)/\rho_{xx}^0$ on the $B^{-1}$ scale, where $\langle\rho_{xx}\rangle$ is the monotonous part of $\rho_{xx}$ and $\rho_{xx}^0$ is $\rho_{xx}(B=0)$.
        (b)~The Fourier spectra of the SdH oscillations.
        (c)~The gate voltage density dependencies for the concentrations obtained from classical measurements ($n_{Drude}$) and the SdH oscillation analysis ($n^{SdH}$). 
        Olive and violet triangles refer to spin-spitted electrons from subband H1, while orange circles -- to electrons from the second subband E2. 
        The sum of all concentrations obtained from the SdH analysis $n_{\Sigma}^{SdH}$ (red squares) fits $n_{Drude}$.
        }
    \end{figure*}

In Fig.~\ref{fig:Shub_and_FFT_diffVg}\,(a) we show the examples of normalized Shubnikov -- de Haas (SdH) resistance oscillations $\Delta\rho_{xx}/\rho_{xx}^0$
measured at different gate voltages vs a reciprocal magnetic field $B^{-1}$.
\textcolor{black}{The corresponding} fast Fourier transform (FFT) spectra are shown in panel~(b).
Several peaks at frequencies $f_i$ occur on each spectrum.
It is seen that at $V_g < 3$\,V there is either one peak $f_1$ or there are two close peaks $f_{1,2}$ that arise on the place of peak $f_1$ at higher densities.
At $V_g > 3$\,V there is also one lower-frequency (e.g., low-density) peak $f_3$.

The frequencies $f_{1,2,3}$ are translated to the density values according to the standard relation $n^\text{SdH}_i = g_i(e/h)f_i$ (where $g_i$ is a spin degeneracy, $e$ is the elementary charge, $h$ is the Planck constant), and shown in Fig.~\ref{fig:Shub_and_FFT_diffVg}\,(c) as empty symbols. 
We assume that $f_1$ and $f_2$ peaks come from spin-resolved Landau levels, so they have $g_{1,2}=1$, and they originate from the first conduction subband H1~(see the inset of Fig.~\ref{fig:Rxx,Ryy_Fridge}\,(a)).
And with an increase of the gate voltage the asymmetry of the QW increases which leads to a strong Rashba spin-orbit splitting\cite{Novik2005} and, hence, distancing of the $f_{1}$ and $f_{2}$ peaks.
This gate voltage asymmetry is enhanced at the high gate voltages due to the natural separation of wave functions with different spin direction to the opposite interfaces\cite{Dobretsova2015a}.
The low-frequency $f_3$ peak is attributed to the spin-degenerate electrons from the second conduction subband E2. 
To prove the suggested carrier identification we plot in Fig.~\ref{fig:Shub_and_FFT_diffVg}\,(c) the total SdH density $n^\text{SdH}_{\Sigma} = n^\text{SdH}_1 + n^\text{SdH}_2 + n^\text{SdH}_3$ (red filled squares) and the density obtained from the Hall measurements (blue line from Fig.~\ref{fig:Rxx,Ryy_Fridge}\,(c)) that coincide quite well.

Furthermore, electrons from the second conduction subband have been recently detected at a similar total density in the cyclotron resonance measurements in Ref.~\onlinecite{Gospodaric2021}, where the same 22-nm HgTe QWs were studied.
This also confirms that the observed low-frequency peak $f_3$ is not related to the magneto-intersubband oscillations (MISO) seen recently in other HgTe-based systems\cite{Minkov2019a}.

Now let us discuss the origin of the resistance anisotropy. 
In fact, there are two types of behaviour on the $\rho_{[03\overline{1}]}/\rho_{[100]}(V_g)$ dependence in Fig.~\ref{fig:Rxx,Ryy_Fridge}\,(b).
First, there is a slow trend of the anisotropy increase with the electron density increase.
Second, there is a sharp anisotropy increase around 2.5\,V as the Fermi level reaches the second subband E2.
Since the electron energy anisotropy in the conduction bands of HgTe~(013) quantum wells is rather weak, the observed resistance anisotropy has to be induced by the scattering anisotropy. 
At low temperatures, when the phonon scattering is frozen, the main scattering sources are the impurities and roughness of QW interfaces. 

\begin{figure}
    \includegraphics[width=0.45\textwidth]{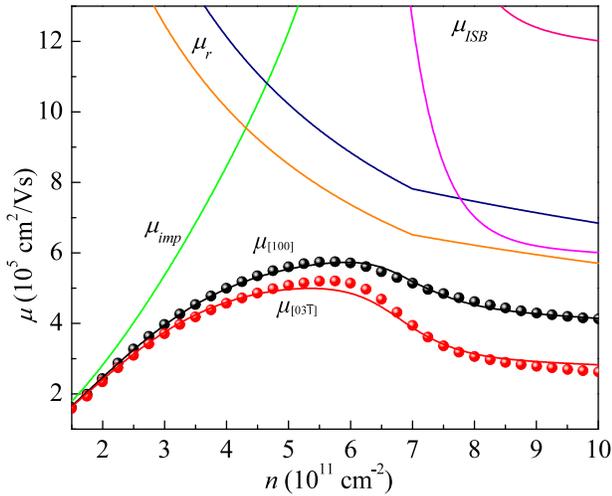}
    \caption{\label{fig:Mu} 
    Experimental (circles) and theoretical (lines) mobilities of electrons in both directions depending on electron concentration $n$. 
    The green line refers to the impurity scattering contribution, blue and orange lines --- to the intra-subband roughness scattering contribution, violet and purple lines --- to the inter-subband roughness scattering contribution for both directions.   
    }
\end{figure}

Here we suggest a simple model to explain the observed anisotropy behaviour. 
Let us assume that only electrons from the first subband H1 contribute to the transport (having higher density and mobility), while E2 electrons participate only in screening and, what is most important, in the inter-subband scattering of H1 electrons (we also neglect the influence of the edge channels that are caused by the inverted spectrum of HgTe \cite{Konig2007} since they have much lower density of states and conductivity in comparison to 2D carriers of the system).
Thus, we consider the following three scattering mechanisms: impurities scattering, intra-subband roughness scattering and inter-subband roughness scattering.
And only the last two are considered to be anisotropic. 
The inter-subband scattering is caused by interface roughness in our model because this mechanism is the only anisotropic one.
For a longitudinal term of the inverted mobility tensor, depending on the in-plane angle $\theta$ from the direction [001], we obtain:
    \begin{equation}
        \frac{1}{\mu} = 
        \frac{1}{\mu}_{imp} + 
        \frac{1+A_r \sin^2\theta}{\mu_{r}} + 
        \frac{1+A_{ISB} \sin^2\theta}{\mu_{ISB}},
    \end{equation}
where $\mu_{imp}$, $\mu_{r}$ and $\mu_{ISB}$ are isotropic contributions of impurities, roughness and inter-subband scattering mechanisms, respectively. 
Here we introduce two phenomenological parameters $A_r$ and $A_{ISB}$ that determine the anisotropy of roughness and inter-subband scattering rates. 
The two steps fitting was performed to determine these parameters~(Fig.\,\ref{fig:Mu}). 
By fitting $\mu(n)$ in the [001] direction with only isotropic contributions we derived the impurity scattering parameters: the number of impurities, height and correlation length of roughness and the intensity of inter-subband scattering
(here we used a model developed in Ref.~\onlinecite{Dobretsova2015a}).
Then, by fitting $\mu(n)$ in the direction $[03\overline{1}]$ we derived $A_r \approx 0.2$ and $A_{ISB} \approx 1$.

Thereby, our phenomenological model shows that, as usual, the impurity scattering dominates at lower electron density values, and the roughness scattering dominates at higher density values. 
So, at lower densities, the anisotropy ratio is closer to unity because the isotropic scattering mechanism, i.e. the impurity scattering,dominates. 
And, at higher densities (but still at $V_{g} < 2V$), the resistance anisotropy increases to 1.1 due to the increase of the impact of the roughness scattering anisotropy with $A_r \approx 0.2$.
And, finally, at even higher density, when E2 subband states appear, the anisotropic inter-subband scattering with $A_{ISB} \approx 1$ starts to dramatically increase the anisotropy ratio.
We stress also that the inter-subband scattering for the non-symmetrical $[03\overline{1}]$ direction turned out to be of the same order as the interface roughness scattering (see $\mu_{r}$ and $\mu_{ISB}$ curves in Fig.\,\ref{fig:Mu}).
And more surprisingly, we found that the inter-subband scattering anisotropy is about 5 times stronger than just roughness scattering anisotropy.

We need to note that it is worth nothing to make a microscopical model that will explain the measured anisotropy in the only first subband by the non-symmetrical  shape of roughness\cite{Markus1994}. 
The problem here is to explain the anisotropy of both inter- and intra-subband scattering together withing such a model. 
We tried to perform such microscopical calculations with different shapes of roughness, but neither elliptical, nor even one-dimensional roughness give such strong anisotropic inter-subband scattering with a simultaneously small anisotropy in the intra-subband scattering. 
The only successful attempt to fit the experiment was to consider that one component of the H1 subband wave function near the interface strongly depends on the in-plane angle, and that only this component scatters to E2 subband states.
Nevertheless, the $kp$ calculations do not support this assumption.

To conclude, we have studied transport anisotropy of electrons in 22-nm wide HgTe\,(013) quantum wells in the directions [001] and $[03\overline{1}]$. 
There is a rather weak anisotropy (up to 10\%) when only one subband H1 is occupied.
And when the Fermi level reaches the second electron subband E2, the anisotropy sharply increases up to 60\%. 
While the first effect can be described by the anisotropic scattering on the nonsymmetrical roughness, such a standard model fails to explain observed both weak and strong anisotropy behaviors withing one set of parameters.
Thus, the microscopic origin of the strong inter-subband anisotropy in HgTe\,(013) systems remains open as a challenge for the theory.
And, more generally, the influence of the roughness anisotropy on the inter-subband scattering and its anisotropy requires more studies in both theory and experiment.

\begin{acknowledgments}
We thank E.\,G.\,Novik, S.\,A.\,Tarasenko and M.\,V.\,Entin for helpful discussions. 

The authors have no conflicts to disclose.
\end{acknowledgments}

\section*{Data Availability Statement} 
The data that support the findings of this study are available from the corresponding author upon reasonable request. 

\bibliography{library}

\end{document}